\documentclass[fleqn,usenatbib]{mnras}

\usepackage{newtxtext,newtxmath}

\usepackage[T1]{fontenc}

\DeclareRobustCommand{\VAN}[3]{#2}
\let\VANthebibliography\thebibliography
\def\thebibliography{\DeclareRobustCommand{\VAN}[3]{##3}\VANthebibliography}

\usepackage{graphicx}	
\usepackage{amsmath}	
\usepackage{color}
\usepackage{enumitem}
\usepackage{hyperref}
\usepackage{verbatim}
\usepackage{amsmath,amstext,mathrsfs}
\usepackage[all]{hypcap} 
\usepackage{afterpage}    
\usepackage{marginnote}
\usepackage{makecell}
\usepackage{subfigure}
\usepackage{graphicx}
\usepackage{epsfig}
\usepackage{natbib}
\usepackage{caption}
\usepackage{multicol,wrapfig}
\usepackage{soul}

\title[Velocity gradient in L1478 and L1482]{Parallel and Perpendicular Alignments of Velocity Gradient and Magnetic Field observed in the Molecular Clouds L1478 and L1482}

\author[Schmaltz et al.]{
Tyler Schmaltz,$^{1,2}$
Yue Hu,$^{1,2}$\thanks{E-mail: yue.hu@wisc.edu}
Alex Lazarian$^{1,2,3}$\thanks{E-mail:alazarian@facstaff.wisc.edu }
\\
$^{1}$Department of Physics, University of Wisconsin-Madison, Madison, WI, 53706, USA\\
$^{2}$Department of Astronomy, University of Wisconsin-Madison, Madison, WI, 53706, USA\\
$^{3}$Centro de Investigación en Astronomía, Universidad Bernardo O’Higgins, Santiago, General Gana 1760, 8370993, Chile
}

\date{Accepted XXX. Received YYY; in original form ZZZ}

\pubyear{2015}

\begin{document}
\label{firstpage}
\pagerange{\pageref{firstpage}--\pageref{lastpage}}
\maketitle

\begin{abstract}
Star formation is a complex process that typically occurs in dense regions of molecular clouds mainly regulated by magnetic fields, magnetohydrodynamic (MHD) turbulence, and self-gravity. However, it remains a challenging endeavor to trace the magnetic field and determine regions of gravitational collapse where the star is forming. Based on the anisotropic properties of MHD turbulence, a new technique termed Velocity Gradient Technique (VGT) has been proposed to address these challenges. In this study, we apply the VGT to two regions of the giant California Molecular Cloud (CMC), namely, L1478 and L1482, and analyze the difference in their physical properties. We use the $^{12}$CO (J = 2 - 1), $^{13}$CO (J = 2 - 1), and C$^{18}$O (J = 2 - 1) emission lines observed with the Heinrich Hertz Submillimeter Telescope. We compare VGT results calculated in the resolutions of $3.3'$ and $10'$ to Planck polarization at 353 GHz and $10'$ to determine areas of MHD turbulence dominance and self-gravity dominance. We show that the resolution difference can introduce misalignment between the two measurements. We find the VGT-measured magnetic fields globally agree with that from Planck in L1478 suggesting self-gravity's effect is insignificant. The best agreement appears in VGT-$^{12}$CO. As for L1482, the VGT measurements are statistically perpendicular to the Planck polarization indicating the dominance of self-gravity. This perpendicular alignment is more significant in  VGT-$^{13}$CO and VGT-C$^{18}$O.
\end{abstract}




\section{Introduction}

Magnetic field and turbulence are pervasive in the interstellar medium (ISM; \citealt{1995ApJ...443..209A,2010ApJ...710..853C,Crutcher12,2014A&A...561A..24B,2017ARA&A..55..111H,HYL20,2022MNRAS.511..829H}). They play a vital role in many astrophysical processes including star formation, molecular cloud evolution \citep{MK04,Crutcher04,MO07,2012ApJ...761..156F,2012ApJ...757..154L}, cosmic ray propagation \citep{1966ApJ...146..480J,2011ApJ...741...16G,2013ApJ...779..140X,Shukurov_2017,2021arXiv211115066H,2022ApJ...934..136X}, and solar flares \citep{2001ApJ...552..833M,Judge_2021,2022LRSP...19....1P}. Nevertheless, magnetic field's role in molecular cloud evolution and star formation is hotly debated, mostly due to the insufficient information on magnetic field structure and strength. Specifically, magnetic fields are challenging to trace and map. Identifying regions where gravity takes over and distinguishing them from regions where gravity is successfully counteracted by the magnetic field and turbulence remain another obscured \citep{Crutcher12}. 


Important efforts to map magnetic fields have been utilized include dust polarization from background starlight \citep{Heiles_2000,2014A&A...561A..24B} or polarized thermal dust emission \citep{BG15,refId0,2019ApJ...872..187C,2021ApJ...913...85H,2022arXiv220512084T}. These are  widely used to trace the plane-of-the-sky (POS) magnetic field orientation in molecular clouds. Zeeman Splitting \citep{Crutcher04,Crutcher12} and Faraday rotation \citep{2019A&A...632A..68T,2022A&A...660A..97T}, on the other hand, can measure the magnetic field strength along the line of sight (LOS). In terms of the POS magnetic field, polarization measurements have considerably advanced our knowledge of the magnetic field in molecular clouds. Nevertheless, the adequate knowledge of the POS magnetic field's variation along the LOS or as a function of gas volume density is still missing. The latter is particularly crucial in understanding star formation as the self-gravity's significance increases in dense regions and it is important to understand at which densities the gravity takes control of the gas dynamics \citep{Crutcher12,2020ApJ...899..115X,HLY20}.
\\

To access the POS magnetic field's variation relative to gas density, we employ the novel Velocity Gradient Technique (VGT; \citealt{Gonz_lez_Casanova_2017,Yuen_2017,Lazarian_2018,10.1093/mnras/sty1807}) to trace the magnetic field's orientation within molecular clouds. VGT works by utilizing the anisotropy of magnetohydrodynamic (MHD) turbulence \citep{1995ApJ...438..763G} that together with the turbulent reconnection \citep{Lazarian_1999} makes turbulent eddies aligned and elongated along the direction of magnetic field that percolates the eddies. Such eddies induce velocity fluctuations that are maximal in the direction perpendicular to the magnetic field at the position of these eddies, this fact supported by numerous simulations \citep{2000ApJ...539..273C,2003MNRAS.345..325C,HXL21}. Consequently, the gradient of velocity fluctuations are maximal when  perpendicular to the magnetic field component revealing this component.   

The turbulence's property, i.e., velocity fluctuation, can be accessed via the Doppler-shifted spectroscopic emission line. Particularly, the emission lines from different molecular species arise at different optical depths corresponding to different gas densities. For instance, $^{12}$CO typically traces the volume density \footnote{Note in this paper, the volume density value traced by molecular species refers to the critical density for excitation by collisions with H$_2$ molecules. The values are adopted from \cite{1978ApJ...222..881G,1998AJ....116..336N,2009ApJ...705L..95I}.} around $10^2~{\rm cm^{-3}}$, while denser C$^{18}$O traces the density roughly around $10^4~{\rm cm^{-3}}$. Applying VGT to various emission lines, therefore, can reveal the magnetic field's variation from diffuse regions to dense regions. This method has been applied with success to multiple different molecular clouds \citep{Hu2019,Hu_2021a,2022MNRAS.510.4952L,2022A&A...658A..90A,2022ApJ...934...45Z}. 

The direction of velocity gradients changes in the regions of gravitational collapse. Thus, VGT can reveal the relative importance of MHD turbulence and self-gravity. Simulations  \citep{2018ApJ...853...96L, HLY20} testify that when the self-gravity dominates over turbulence the velocity gradients are dominated by the inward collapse parallel to the magnetic field. This change is characterized by a shift in the cloud dynamics where the VGT orientation flips by at most 90 degrees to align parallel with the magnetic field inferred from polarization \citep{HLY20}. Such a change of direction observed in a given emission line suggests the volume density threshold that can trigger the gravitational collapse.

The VGT has been proven to be a reliable tool for tracing the magnetic field in diffuse media and in clouds with low star formation rate \citep{HLY20,2020MNRAS.496.2868L,Hu_2021b}. In this study we explore the performance of VGT for the massive star-forming regions. For our research, we chose to study the L1478 and L1482 star-forming regions of the California molecular cloud (CMC), which is roughly 450 pc \citep{2009ApJ...703...52L} away from Earth. L1478 and L1482 are both chosen for their distinguishable properties. Although they are two neighboring filamentary regions, their orientation relative to the Galactic plane is different, i.e., L1478 orientates to the northwest while L1482 is northeast \citep{Lewis_2021}. Particularly, L1478 has a lower column density and is more diffuse than L1482 suggesting the significance of self-gravity within the two regions might be different \citep{2019ApJ...877..114C}. These facts make these two regions prime targets for studying the magnetic field and self-gravity's effects on molecular cloud's evolution through VGT.

In order to calculate the velocity gradient, we used $^{12}$CO (J = 2 - 1), $^{13}$CO (J = 2 - 1), and C$^{18}$O (J = 2 - 1) spectral lines observed with the Heinrich Hertz Submillimeter Telescope (SMT) \citep{Lewis_2021}. These three molecules exhibit different optical depths and are sensitive to different density ranges. $^{12}$CO typically is optically thick (optical depths $\gg1$) tracing the cloud's outskirt diffuse region, while 
C$^{18}$O is optically thin (optical depths $\ll1$) imprinting the information of the cloud's dense core \citep{Lewis_2021,2019ApJ...884..137H}. This fact allows us to form a 3D magnetic field tomography within the molecular cloud along the LOS utilizing VGT. Additionally, to investigate the significance of self-gravity, we compare the VGT-derived magnetic fields with the one inferred from the Planck dust polarization at 353 GHz \citep{2020A&A...641A...3P}. 



In this research, we measure the magnetic field of L1478 and L1482 regions of the California molecular cloud with the VGT method. This paper is organized into six sections. In \S~\ref{sec:data}, we describe the observational data that was used for this research. In \S~\ref{sec:method}, we introduce the methodology behind VGT and its established pipeline. In \S~\ref{sec:results}, we present our observational results of magnetic fields traced by VGT using different tracers and then compare our results with Planck polarization results. In \S~\ref{sec:dis}, we discuss the importance of VGT and its applicability. We also discuss its agreement and anti-agreement with Planck polarization and its physical implication for the CMC. Finally, in \S~\ref{sec:con}, we conclude with a summary of our work.






\begin{table*}
	\centering
	\caption{This table displays the mean AM values for the corresponding 60$\times$60 (i.e., AM$_{60}$) or 20$\times$20 (i.e., AM$_{20}$) sub-block sizes with corresponding standard deviations for both L1478 and L1482 regions of the CMC. Their corresponding uncertainties $\sigma_{\rm AM_{60}}$ and $\sigma_{\rm AM_{20}}$ are given by the standard deviation of the mean. }
	\label{tab:example_table}
	\begin{tabular}{|c|c|c|c|c|c|} 
		\hline
		Cloud & Emission line & Mean AM$_{20}$ & Mean AM$_{60}$ & $\sigma_{\rm AM_{20}}$ & $\sigma_{\rm AM_{60}}$ \\
		\hline
		L1478 & $^{12}$CO & 0.56 & 0.70 & 0.002 & 0.001\\
		 & $^{13}$CO & 0.31 & 0.42 & 0.002 & 0.002\\
		 & C$^{18}$O & 0.59 & 0.49 & 0.003 & 0.003\\
		\hline\hline
		L1482 & $^{12}$CO & 0.06 & 0.02 & 0.003 & 0.003\\
		 & $^{13}$CO & -0.22 & -0.40 & 0.002 & 0.002\\
		 & C$^{18}$O & -0.29 & -0.45 & 0.003 & 0.002\\
		\hline
	\end{tabular}
\end{table*}

\section{Observational data}
\label{sec:data}
\subsection{Emission line}
We used data from \cite{Lewis_2021} whom observed $^{12}$CO (J = 2 - 1), $^{13}$CO (J = 2 - 1), and C$^{18}$O (J = 2 - 1) spectral line measurements of L1478 and L1482 from the Heinrich Hertz Submillimeter Telescope using a prototype ALMA Band 6 dual-polarization sideband separating receiver in combination with the 0.25 MHz—256 channel filterbank as the back end (0.25 MHz $\sim0.32$ km s$^{-1}$ at 230 GHz). The beam and velocity resolution for each CO tracer were slightly different, the maps were therefore convolved to the same resolution of the $^{12}$CO (38") and regridded into pixel resolution 10" and velocity resolution to 0.3 km s$^{-1}$ to allow for comparison. As for noise level, the average rms noise per 0.3 km s$^{-1}$ channel was 0.005 K for C$^{18}$O, 0.03 K for $^{13}$CO, and 0.07 K for $^{12}$CO. The averaged spectrum for each line is presented in Appendix B.

To derive the column density map, we adopt the X-factor method used in \cite{2008ApJ...679..481P}. The X-factor of $^{12}$CO, i.e., $X_{12}$ is defined as:
\begin{equation}
X_{12}=W_{12}/N_{\rm H_2},    
\end{equation}
where $W_{12}$ is the integrated $^{12}$CO intensity and $N_{\rm H_2}$ is the column density map. For simplicity, we use a uniform value $X_{12}=2\times10^{20}~{\rm cm^{-2}K^{-1} km^{-1} s}$ \citep{2008ApJ...679..481P}. To derive column densities from the $^{13}$CO and C$^{18}$O intensity maps, we follow \citep{2008ApJ...679..481P}:
\begin{equation}
\begin{aligned}
    N_{\rm H_2}&=A_v(9.4\times10^{20}~{\rm cm^{-2} mag^{-1}}),\\ 
    A_v&=
    \begin{cases}
    W_{13}(0.345~{\rm mag^{-1} K^{-1} km^{-1} s})+1.46~{\rm mag^{-1}},~{\rm for~^{13}CO}\\
    W_{18}(2.9~{\rm mag^{-1} K^{-1} km^{-1} s})+2.4~{\rm mag^{-1}},~{\rm for~C^{18}O}\\
    \end{cases},
\end{aligned}
\end{equation}
where $A_v$ is the extinction. $W_{13}$ and $W_{18}$ are integrated $^{13}$CO and C$^{18}$O intensities, respectively. 

\subsection{Planck polarization}
In our research we compared the magnetic field inferred from velocity gradient to the one obtained from Planck polarization data \citep{2020A&A...641A...3P}. We use the Planck 353 GHz polarized dust signal data from the Planck 3rd Public Data Release (DR3) 2018 of High Frequency Instrument. Observations from Planck designate the polarization angle $\phi$ with Stokes parameter maps $I$, $Q$, and $U$. $\phi$ is defined mathematically as:
\begin{equation}
\phi=\frac{1}{2}\tan^{-1}(-U,Q),
\end{equation}
where the -U notation converts the angle from HEALPix convention to IAU convention. The Stokes parameter maps were smoothed from nominal angular resolution 5$'$ to 10$'$ with a Gaussian kernel to suppress noise and achieve a higher sign-to-noise ratio. We infer the magnetic field angle from the equation: $\phi_B$ = $\phi$ + $\pi$/2.

Note that the foreground contribution is not calibrated in the Planck data. L1478 and L1482 are nearby clouds with an infrared extinction derived from Herschel's observations 1.5 - 4 \citep{2017A&A...606A.100L}. The extinction is much larger than that of the diffuse foreground. We, therefore, expect the Planck polarization to be dominated by the signal from the clouds.

\begin{figure*}
	\includegraphics[width=0.75\linewidth]{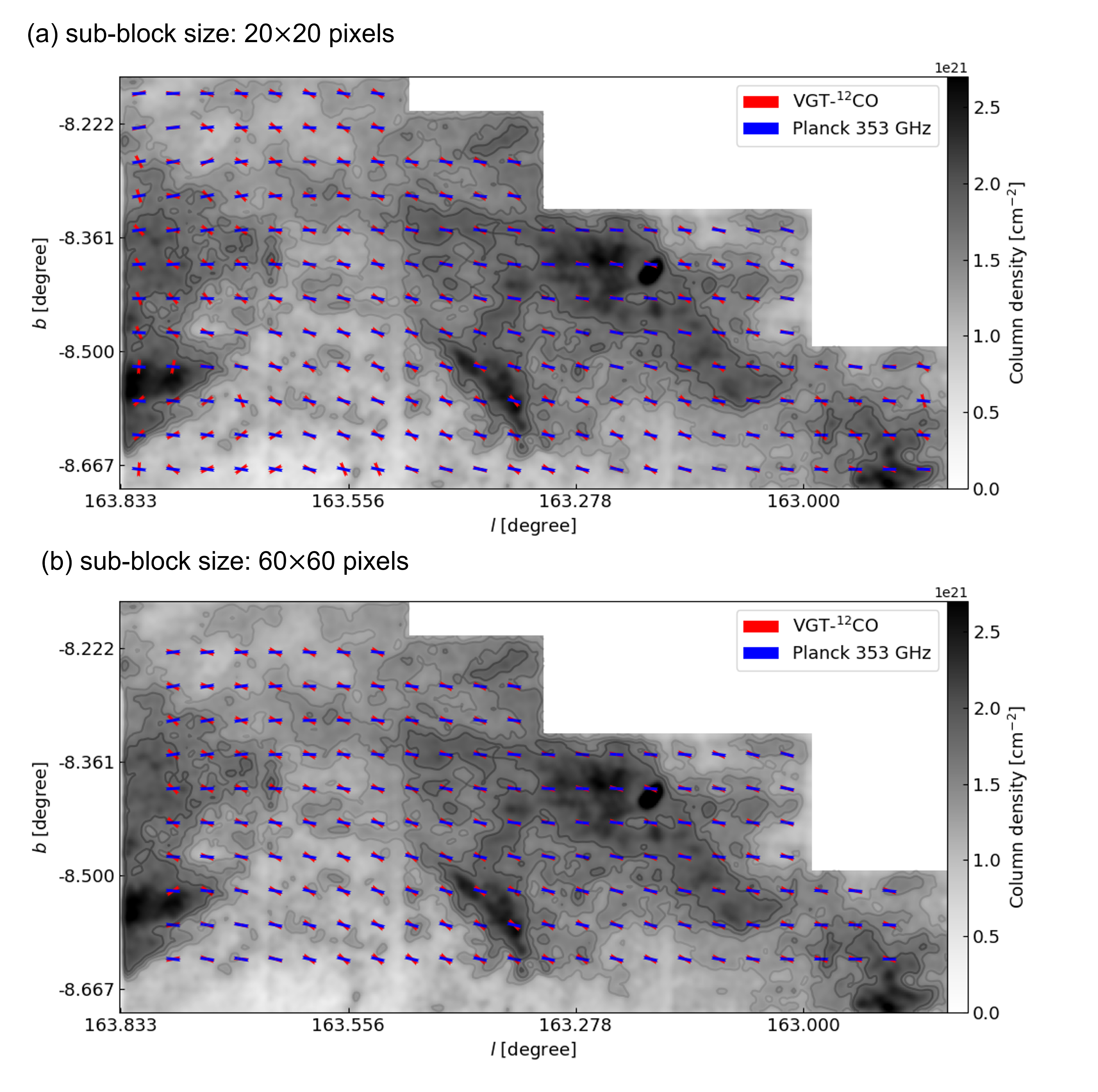}
    \caption{The magnetic field orientation inferred from the VGT using $^{12}$CO for sub block sizes of 20$\times$20 pixels ($\sim0.4$~pc; top) and 60$\times$60 pixels ($\sim1.2$~pc; bottom) in the L1478 region of the CMC. VGT (red) and Planck polarization (blue) segments are overlaid in the integrated $^{12}$CO intensity map. Contours of both maps approximately at $1.3\times10^{21}$,  $1.4\times10^{21}$, $1.6\times10^{21}$, and $1.8\times10^{21}$ cm$^{-2}$.}
    \label{fig:L1478 12CO dn=20 and dn=60 VGT vs Planck}
\end{figure*}

\begin{figure*}
	\includegraphics[width=0.7\linewidth]{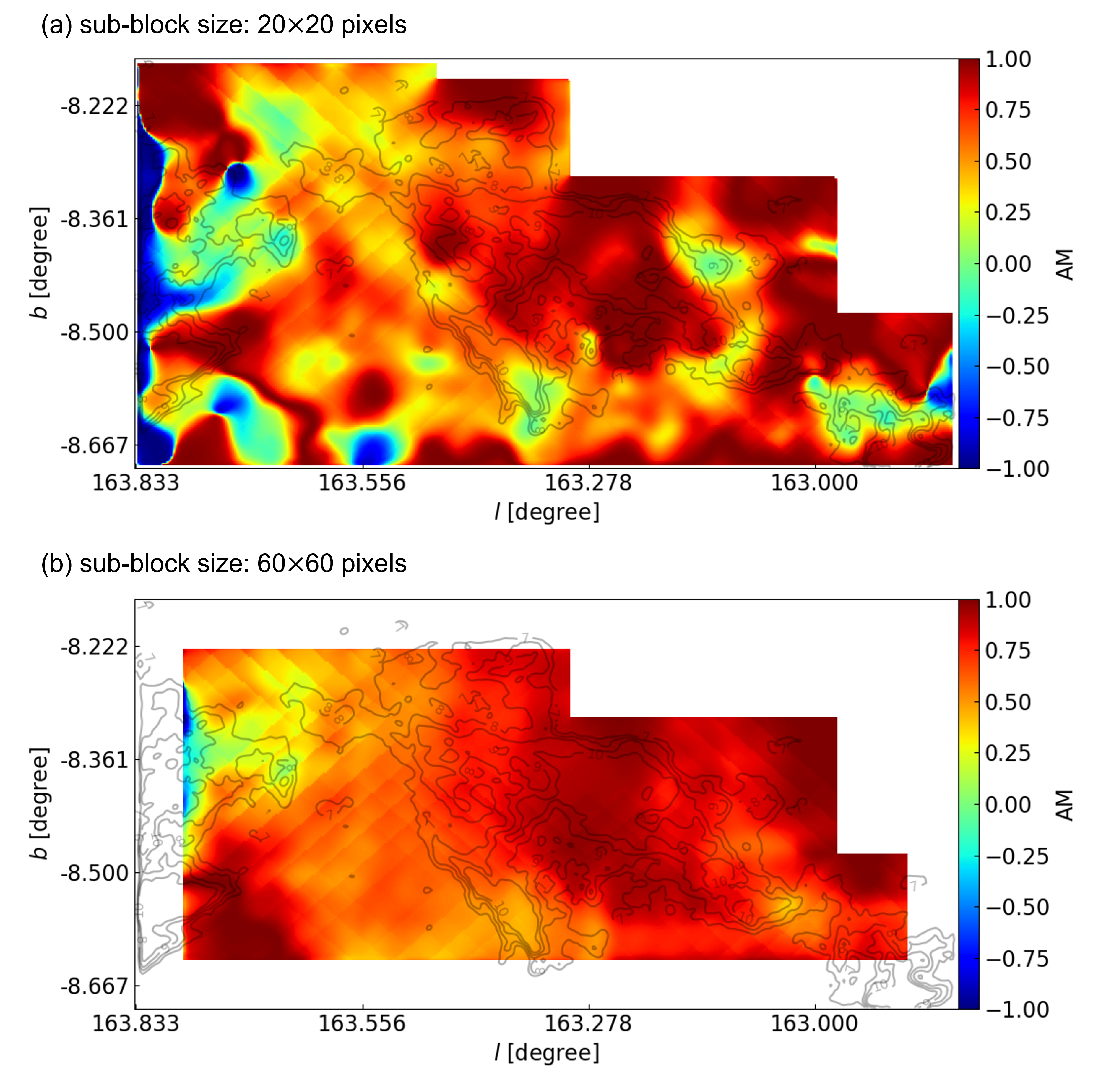}
    \caption{The distribution of AM calculated from the Planck and VGT measurements using $^{12}$CO for sub block sizes of 20$\times$20 pixels ($\sim0.4$~pc; top) and 60$\times$60 pixels ($\sim1.2$~pc; bottom) in the L1478 region of the CMC. Red regions represent parallel alignment (i.e., AM values of positive one) and blue regions represent perpendicular alignment (i.e., AM values of negative one). Contours of both maps approximately at $1.3\times10^{21}$,  $1.4\times10^{21}$, $1.6\times10^{21}$, and $1.8\times10^{21}$ cm$^{-2}$.}
    \label{fig:L1478 12CO dn=20 and dn=60 AM Map}
\end{figure*}

\begin{figure*}
	\includegraphics[width=0.85\linewidth]{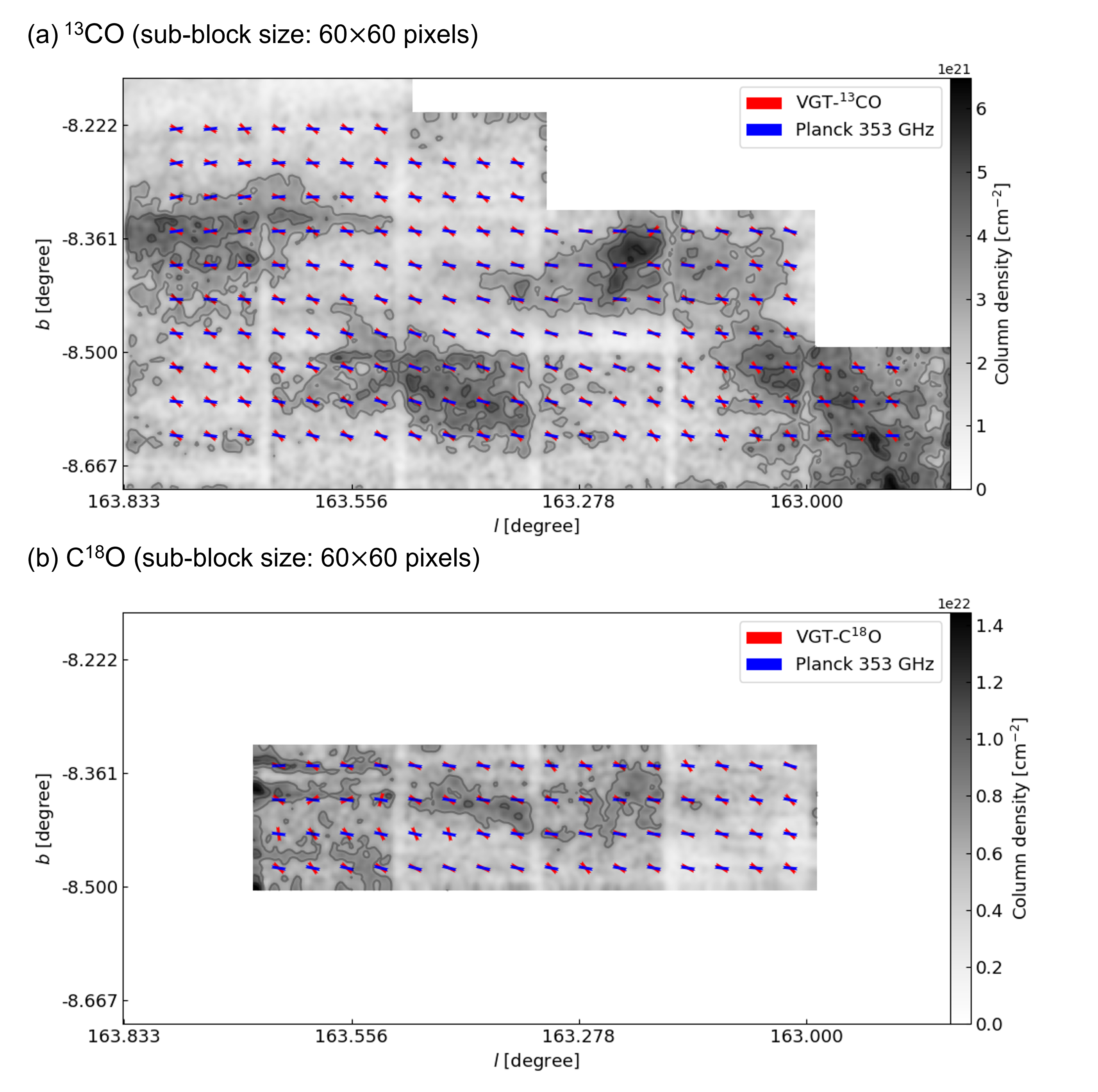}
    \caption{The magnetic field orientation inferred from the VGT using $^{13}$CO (top) and C$^{18}$O (bottom) with sub block sizes of 60$\times$60 pixels ($\sim0.4$~pc) in the L1478 region of the CMC. VGT gradient lines are represented in red and Planck polarization lines are represented in blue. Contours for top map are at $2.1\times10^{21}$, $2.8\times10^{21}$, $3.5\times10^{21}$, and $4.1\times10^{21}$ cm$^{-2}$ and for bottom map are $6.6\times10^{21}$, $8.8\times10^{21}$, $1.1\times10^{22}$, and $1.3\times10^{22}$ cm$^{-2}$.}
    \label{fig:L1478 13CO and C18O dn = 60 VGT vs Planck}
\end{figure*}

\begin{figure*}
	\includegraphics[width=0.8\linewidth]{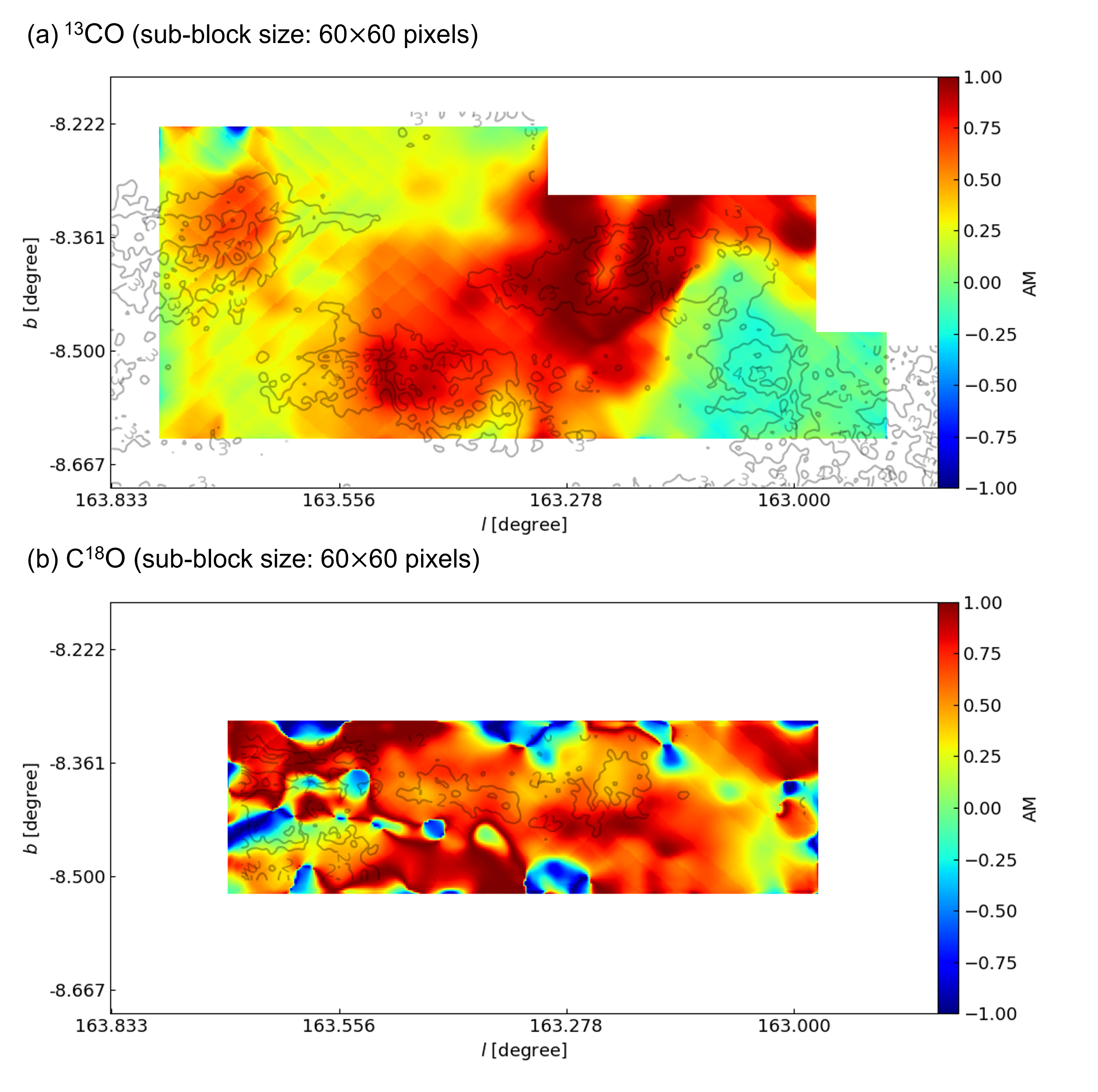}
    \caption{The AM distribution calculated from the Planck and VGT gradients using $^{13}$CO (top) and C$^{18}$O (bottom) with sub block sizes of 60$\times$60 pixels ($\sim0.4$~pc) in the L1478 region of the CMC. Contours for top map are at $2.1\times10^{21}$, $2.8\times10^{21}$, $3.5\times10^{21}$, and $4.1\times10^{21}$ cm$^{-2}$ and for bottom map are $6.6\times10^{21}$, $8.8\times10^{21}$, $1.1\times10^{22}$, and $1.3\times10^{22}$ cm$^{-2}$.}
    \label{fig:L1478 13CO and C18O dn = 60 AM Map}
\end{figure*}

\section{Methodology}
\label{sec:method}
\subsection{Basics of VGT}
VGT is theoretically based on the theories of MHD turbulence \citep{1995ApJ...438..763G} and fast turbulent reconnection \citep{Lazarian_1999}. It has been applied to study magnetic fields in both diffuse and molecular species 
\citep{HLY20,2020MNRAS.496.2868L,Hu_2021b,2022arXiv220512084T,2022MNRAS.511..829H}.  

The scale-dependent anisotropy of magnetic fluctuations was predicted in \cite{1995ApJ...438..763G}:
\begin{equation}
\label{eq.gs95}
\begin{aligned}
          \textit{$k_{\parallel}$}  \propto   \textit{$k_{\perp}^{2/3}$},
\end{aligned}
\end{equation}
where $k_\bot$ and $k_\parallel$ are wavevectors perpendicular and parallel to magnetic field. In the original study, the magnetic field was the mean magnetic field. It is important to mention, that in Fourier space, the local spatial information is not available so that the anisotropy is measured with respect to the mean magnetic field. This anisotropy in fact is dominated by the largest eddy and is scale-independent \citep{2000ApJ...539..273C,HXL21}. 

Nevertheless, turbulent reconnection predicts the existence of eddies that reconnect over the eddy turnover time. Therefore, the magnetic field does not impede the evolution of such eddies and most energy of the cascade is channeled to them. However, it is obvious that the eddies can feel only the magnetic field around them. Thus the parallel and perpendicular scales of eddies $l_\parallel$ and $l_\perp$ should be identified in terms of the local direction of magnetic field. In these variables,
the modified critical balance condition means that the eddy cascading time ($l_{\perp}/v_l$) equals the wave period ($l_{\parallel}/v_{\rm A}$), where $v_{\rm A} $ is the Alfv\'en velocity and $v_l$ is turbulent velocity at the perpendicular scale $l_\perp$. According to \cite{1999ApJ...517..700L}, turbulent eddies, due to the fast magnetic reconnection, mix up the magnetic field in perpendicular direction. The cascade follows the Kolmogorov scaling $v_{l,\bot}\propto l_{\bot}^{1/3}$. Consequently, one can get the relation between the parallel and perpendicular scales of the eddies in the local reference frame (i.e., defined by local magnetic fields, see \citealt{1999ApJ...517..700L}):
 \begin{equation}
\begin{aligned}
l_{\parallel}  =  L_{\rm inj}(\frac{l_{\bot}}{L_{\rm inj}})^{2/3}M_{\rm A}^{-4/3},~M_{\rm A} \le 1,
\end{aligned}
\end{equation}
where $L_{\rm inj}$ is the injection scale and $M_{\rm A}=v_{\rm inj}/v_A$ is the Alfv\'en Mach number, $v_{\rm inj}$ is turbulence's injection velocity. 

Particularly, due to the anisotropy $l_\bot\ll l_\parallel$, the amplitude of velocity fluctuation and its corresponding gradient are \citep{1999ApJ...517..700L}:
\begin{equation}
\label{eq.4}
        \begin{aligned}
           &v_{l,\bot}  =  v_{\rm inj}(\frac{l_{\bot}}{L_{\rm inj}})^{1/3}M_{\rm A}^{1/3},~M_{\rm A} \le 1,\\
           &\nabla v_l\propto \frac{v_{l,\bot}}{l_{\bot}}=\frac{v_{\rm inj}}{L_{\rm inj}}M_{\rm A}^{1/3}(\frac{l_{\bot}}{L_{\rm inj}})^{-2/3},~ M_{\rm A} \le 1,
        \end{aligned}
\end{equation}
which reveal that the velocity gradient $\nabla v_l$ is dominated by its perpendicular component and can trace the 3D direction of magnetic field. Note that the anisotropy appears when the local magnetic field’s role is at least comparable to turbulence, i.e., MHD turbulence. For the $M_{\rm A}\gg1$ case, turbulence is isotropic because it is essentially hydrodynamic turbulence. However, the significance of turbulence is cascading from large injection scales ($L_{\rm inj}\sim100$~pc, see \citealt{2010ApJ...710..853C}) to smaller scales, which means the turbulent velocity decreases. Eventually, at the transition scale $l_a = L_{\rm inj}/M_{\rm A}^3$, magnetic field and turbulence are comparable \citep{2006ApJ...645L..25L}. For L1478 and L1482, their length size on the POS is around $5$~pc. Therefore, if we consider $l_a<5$~pc, the cloud is in the condition of $M_{\rm A}>3$. The typical value for diffuse molecular clouds, however, are sub-Alfv\'enic $M_{\rm A}<1$ or trans-Alfv\'enic $M_{\rm A}\approx1$ \citep{2022arXiv220311179P}. We therefore expect turbulence in L1478 and L1482 to be anisotropic.

The above consideration is valid for MHD turbulence in diffuse ISM, where self-gravity can be ignored. In star forming regions, self-gravity might be significant and dominate over MHD turbulence. In the case of gravitational collapse, i.e., strong self-gravity, the nature of turbulent flow is modified. In the direction perpendicular to the magnetic field, any gravitational pull is counteracted by a magnetic force. The most significant acceleration of the plasma, therefore, appears in the direction parallel to the magnetic field. The acceleration results in huge velocity difference along the magnetic field so that the velocity gradients are expected to change their orientation from perpendicular to magnetic fields to align with magnetic filed. This property has been numerically and observationally confirmed serving as an important tool in identifying gravitational collapse \citep{HLY20,Hu_2021b}.

\subsection{VGT pipeline}
To extract the velocity information in observation, in this research, we use the thin velocity channel Ch($x$,$y$) of emission lines \cite{Lazarian_2018}.The theory of the fluctuations in such channels is formulated in \cite{2000ApJ...537..720L}.  According to the theory the narrower the channel width, the more significant the contribution of velocity fluctuations. For the density spectrum dominated by large scale velocity fluctuations, if the channel width $\Delta v$ is smaller than the velocity dispersion $\sqrt{\delta (v^2)}$ of turbulent eddies under study, i.e., $\Delta v < \sqrt{\delta (v^2)}$, the intensity fluctuation in a thin channel is dominated by velocity fluctuation. Otherwise, the intensity fluctuation is dominated by density fluctuation in a thick channel. Therefore, the intensity gradient calculated in a thin velocity channel map, referred as Velocity Channel Gradients (VChGs) contains the information of velocity fluctuation \citep{Lazarian_2018}. 

Explicitly, the gradient map $\psi_{\rm g}$ can be calculated using these equations:
\begin{equation}
\label{eq.5}
\begin{aligned}
\nabla_x{\rm Ch}_i &= {\rm Ch}_i(x,y) - {\rm Ch}_i(x-1,y),\\
\nabla_y{\rm Ch}_i &= {\rm Ch}_i(x,y) - {\rm Ch}_i(x,y-1),\\
\psi_{g}^{i}  &= {\rm tan}^{-1}(\frac{\nabla_y{\rm Ch}_i(x,y)}{\nabla_x{\rm Ch}_i(x,y)}),
\end{aligned}
\end{equation}
where $\nabla_{x}$Ch$_{i}(x,y)$ and $\nabla_{y}$Ch$_{i}(x,y)$ represent the $x$ and $y$ components of the gradient, respectively. The subscript $i = 1,2,...,n_v$ stands for the $i^{\rm th}$ channel map with total number $n_v$. These equations are applied to all pixels with spectral line emissions having a signal-to-noise ratio greater than 3.

The gradient orientation given in Eq.~\ref{eq.5} for each pixel, however, is not statistically sufficient to derive turbulence's property. The gradient map $\psi_{\rm g}$ is further processed by the sub-block averaging method \citep{Yuen_2017} to extract turbulence's anisotropy. This method takes all velocity gradients orientation within a specific sub-block and plots a corresponding histogram. A Gaussian fitting is applied to the histogram and the gradient orientation corresponding to Gaussian distribution's peak value is taken as the mean gradient for that sub-block (see Fig.~\ref{fig:subblock}). We denoted the processed gradient map as $\psi^{i}_{gs}(x,y)$. The map's edges with pixel numbers less than the sub-blocks size are not considered in the calculation.

To compare with polarization measurement, we construct the Pseudo-Stokes-parameters $Q_g$ and $U_g$ from  $\psi^{i}_{gs}(x,y)$:
\begin{equation}
\begin{aligned}
Q_g(x,y) &= \sum_{i=1}^{n_v} I_i(x,y){\rm cos}(2\psi^{i}_{gs}(x,y)),\\
U_g(x,y) &= \sum_{i=1}^{n_v} I_i(x,y){\rm cos}(2\psi^{i}_{gs}(x,y)),\\
\psi_g &= \frac{1}{2}{\rm tan}^{-1}(\frac{ U_g}{Q_g}),
\end{aligned}
\end{equation}
where $\psi_g$ is the pseudo polarization angle which is perpendicular to the POS magnetic field. The magnetic field orientation is then inferred from
\begin{equation}
    \psi_B = \psi_g + \pi/2.
\end{equation}
Note $\psi_B$ is defined as east through the north, while $\phi_B$ (from Planck polarization) is defined as north through the west. We transform the Planck angle to VGT’s angle coordinate for comparison.

The alignment between the VGT measurement and polarization measurement is quantified by the Alignment Measure (AM; \citealt{Gonz_lez_Casanova_2017}), defined as:
\begin{equation}
        \begin{aligned}
           {\rm AM}  =  2(\cos^2\theta_r-\frac{1}{2}),
        \end{aligned}
\end{equation}
where $\theta_r$ is the relative angle between the two angles. AM is a relative scale ranging from -1 to 1, with AM = 1 indicating that two angles are parallel, i.e., turbulence dominated and AM = -1 denoting that the two are orthogonal, i.e., self-gravity dominated. The averaged AM values for L1478 and L1482 at resolutions 10$'$ and $3.3'$ are listed in Tab.~\ref{tab:example_table}.

\section{Observational Results}
\label{sec:results}
\subsection{L1478}
Fig.~\ref{fig:L1478 12CO dn=20 and dn=60 VGT vs Planck} presents the magnetic field inferred from VGT using the $^{12}$CO emission line. To compare with the magnetic field orientation revealed by Planck polarization, we first choose sub-block size $60\times60$ pixels for the sub-block averaging method. Such a sub-block size gives an effective resolution of $10'$ ($\sim1.2$~pc) which is comparable to the Planck measurement. We can see the two measurements have a good agreement giving mean AM$\sim0.70$ (see Tab.~\ref{tab:example_table}). The spatial distribution of AM is given in Fig.~\ref{fig:L1478 12CO dn=20 and dn=60 AM Map}. Further, we decrease the sub-block size to 20$\times$20 size so that we can trace the magnetic field at a higher resolution.

The VGT-$^{12}$CO-measured magnetic field map, as well as the corresponding AM distribution, at resolution 3.3$'$ ($\sim0.4$~pc) are presented in Figs.~\ref{fig:L1478 12CO dn=20 and dn=60 VGT vs Planck} and \ref{fig:L1478 12CO dn=20 and dn=60 AM Map}. The decrease of the sub-block size reveals smaller scale variations of magnetic field. In the 3.3$'$ map, we see a much more perpendicular alignment of the VGT measurement with Planck polarization than in the 10$'$ map, particularly for the eastern part. The disagreement here is purely caused by the difference in resolution. This suggests that the magnetic field at a smaller scale exhibits variation, which cannot be resolved by Planck polarization.

We further use $^{13}$CO and C$^{18}$O emission lines to map the magnetic field. $^{13}$CO typically traces gas at volume density $\sim10^3~{\rm cm^{-3}}$ and optically thin (optical depth $\ll 1$) C$^{18}$O trace denser gas at density $\sim10^4~{\rm cm^{-3}}$. The application of VGT to the two tracers provides magnetic field information in denser regions. The VGT-$^{13}$CO (or VGT-C$^{18}$O) and Planck polarization has less alignment than the case of VGT-$^{12}$CO. At resolution 10$''$, the mean AM of VGT-$^{13}$CO and VGT-C$^{18}$O is 0.42 and 0.49, respectively (see Tab.~\ref{tab:example_table}). The magnetic field and AM maps at resolution 10$'$ are presented in Figs.~\ref{fig:L1478 13CO and C18O dn = 60 VGT vs Planck} and \ref{fig:L1478 13CO and C18O dn = 60 AM Map}. Compared with VGT-$^{12}$CO at 10' measurement (see Fig.~\ref{fig:L1478 12CO dn=20 and dn=60 AM Map}), we see the appearance of more misalignment in $^{13}$CO and more perpendicular alignment in C$^{18}$O. For $^{13}$CO, misalignment appears mostly in low intensity regions especially for the western and eastern part. In Fig.~\ref{fig:L1478 13CO Histogram} at 10$'$ we see much of the regions with misalignment occurring around the AM value of -0.25. From Tab.~\ref{tab:example_table}), the mean AM value for $^{13}$CO at 10$'$ is 0.42 which points to more misalignment than $^{12}$CO. One possibility for the misalignment is the contribution from the foreground and background dust polarization. It is known (see \citealt{2022MNRAS.510.4952L}), low-intensity regions typically correspond to also low extinction. It is possible that in these low-extinction regions, the atomic H I or molecular $^{12}$CO foreground/background contribute more to the dust polarization. For C$^{18}$O focusing only on a central smaller region, majority of regions have parallel alignment. We can see in Fig.~\ref{fig:L1478 C18O Histogram} at 10$'$, that the histogram spreads and is centered around more positive AM values than negative with a mean AM value of 0.49. In view that $^{12}$CO, $^{13}$CO, and C$^{18}$ traces up to the volume density of $10^4~{\rm cm^{-3}}$, the positive AM at resolution 10$'$ suggests that L1478 globally is dominated by MHD turbulence and the self-gravity's effect is not dominant up to volume density $10^4~{\rm cm^{-3}}$.

While we identify magnetic field using polarization, one must understand the limitations of this approach. Radiative torques (RATs) align grains in the settings where the radiation at the wavelength comparable with the dust size is sufficient \citep{2007MNRAS.378..910L}. This, by itself, makes integration of magnetic field along line of sight inhomogeneous and different from the way that VGT probes magnetic fields. In addition, RATs can destroy grains \citep{2019ApJ...876...13H}. The dust grains aligned at attractor points with high angular momentum \citep{2021ApJ...908...12L}. The rotationally destroyed dust grains become too small to be span up and aligned by RATs. As a result, the sampling of magnetic field by polarization can be inhomogeneous not only in the denser darker regions where RATs are weak, but also in regions of high illumination where RATs are too strong.


\begin{figure}
	\includegraphics[width=0.9\linewidth]{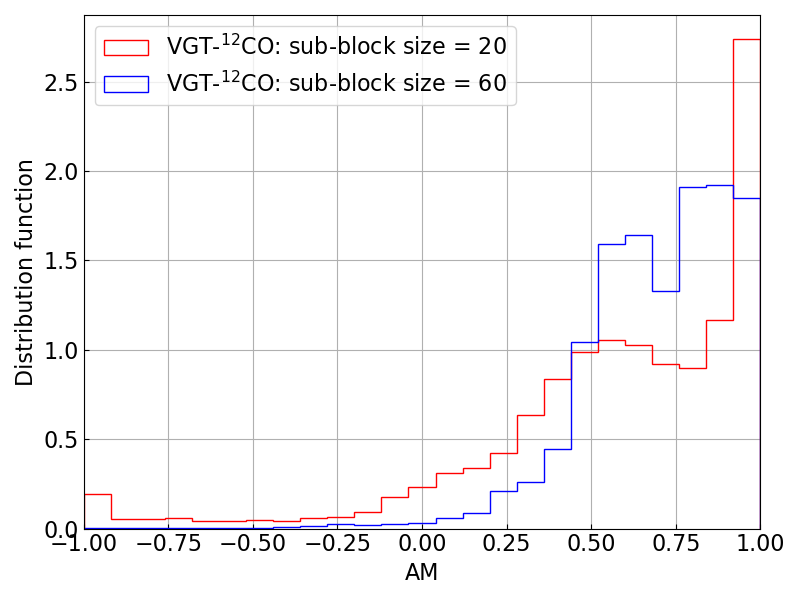}
    \caption{The histogram of AM values inferred from alignment between Planck Polarization and VGT using $^{12}$CO in the L1478 region of the CMC.}
    \label{fig:L1478 12CO Histogram}
\end{figure}

\begin{figure}
	\includegraphics[width=0.9\linewidth]{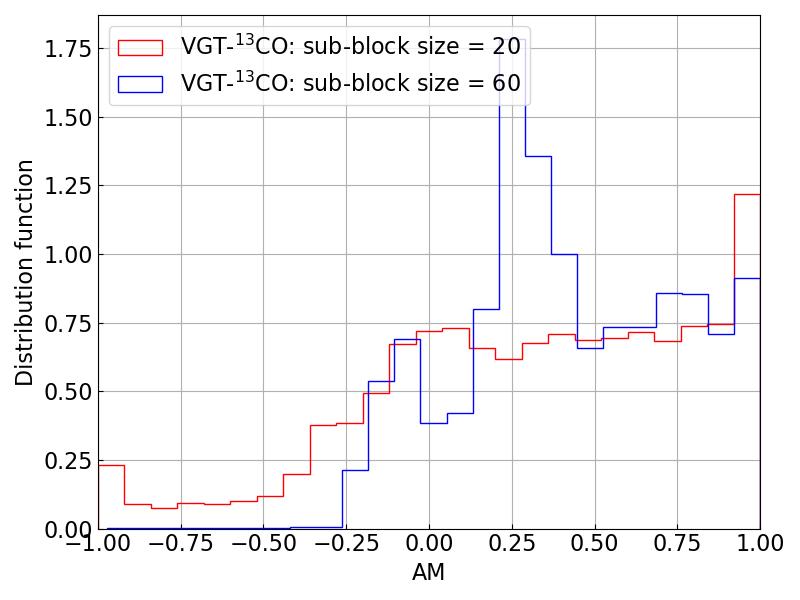}
    \caption{Same as Fig.~\ref{fig:L1478 12CO Histogram}, but for $^{13}$CO.}
    \label{fig:L1478 13CO Histogram}
\end{figure}

\begin{figure}
	\includegraphics[width=0.9\linewidth]{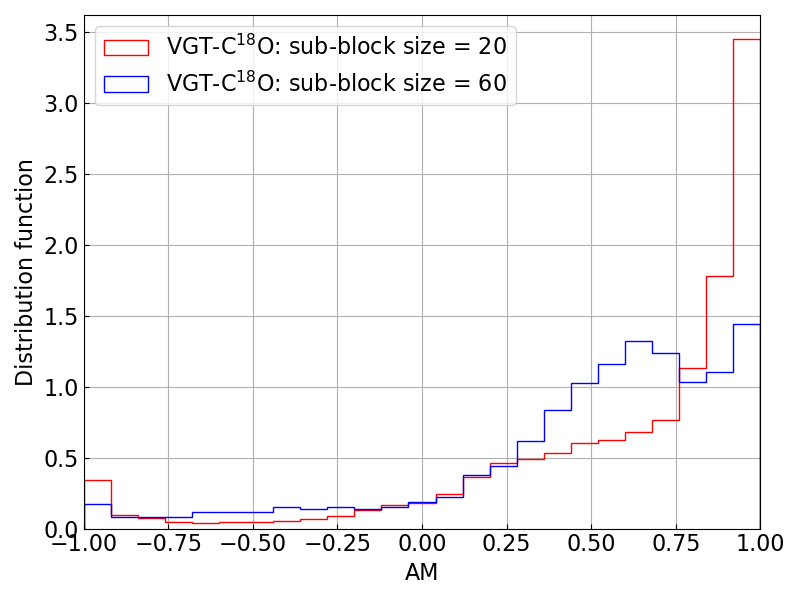}
    \caption{Same as Fig.~\ref{fig:L1478 12CO Histogram}, but for C$^{18}$O.}
    \label{fig:L1478 C18O Histogram}
\end{figure}

\begin{figure*}
	\includegraphics[width=0.99\linewidth]{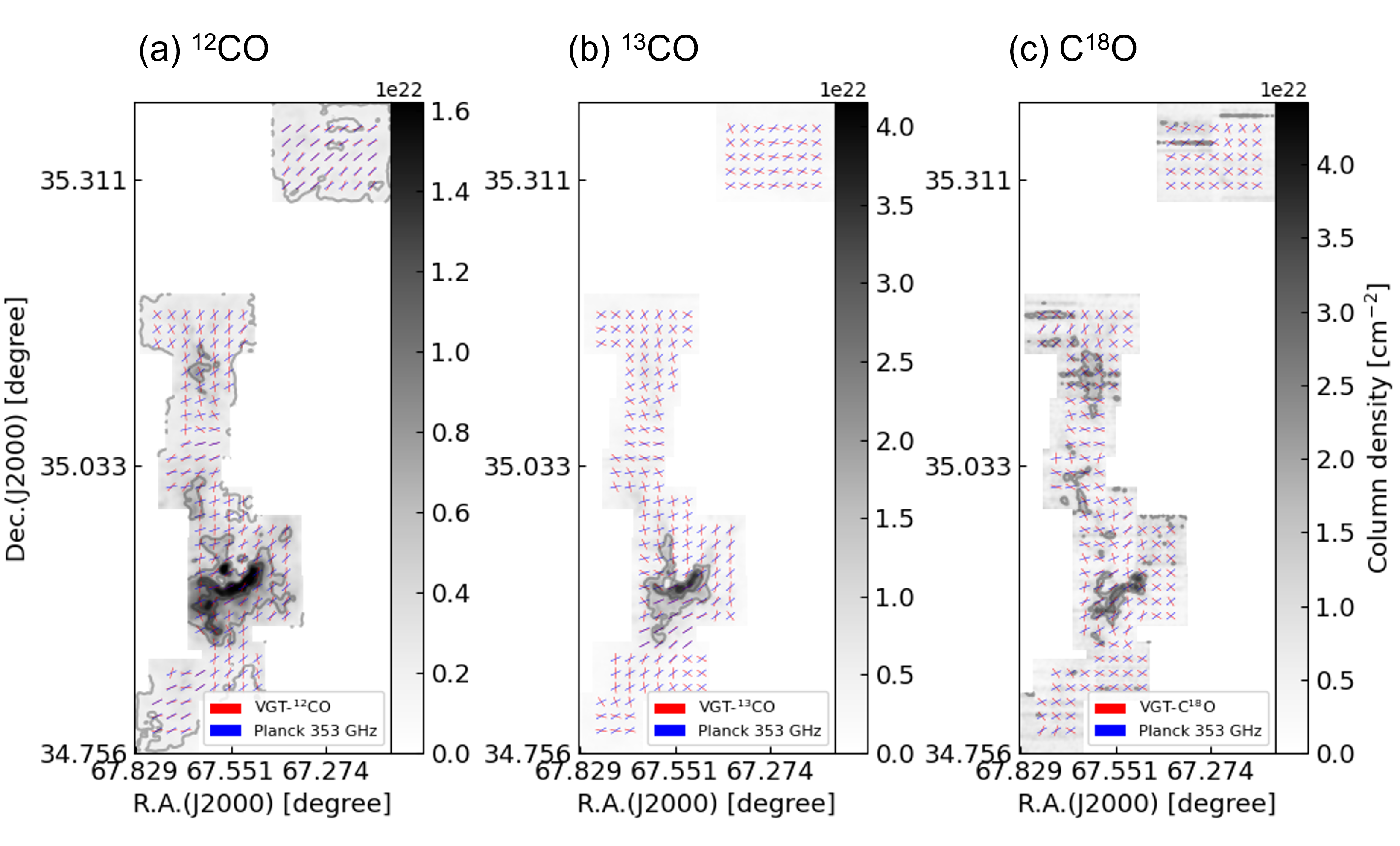}
    \caption{Magnetic field orientation inferred from the VGT using $^{12}$CO (left) with contours at $1.8\times10^{21}$, $5.4\times10^{21}$, $1.1\times10^{22}$, and $1.4\times10^{22}$ cm$^{-2}$, $^{13}$CO (middle) with contours at $1.0\times10^{22}$, $2.1\times10^{22}$, $3.1\times10^{22}$, and $3.8\times10^{22}$ cm$^{-2}$, and C$^{18}$O (right) with contours at $1.3\times10^{22}$, $2.2\times10^{22}$, $3.1\times10^{22}$, and $4.0\times10^{22}$ cm$^{-2}$, respectively for sub block sizes of 60$\times$60 pixels ($\sim1.2$~pc) in the L1482 region of the CMC. VGT gradient lines are represented in red and Planck polarization lines are represented in blue.}
    \label{fig:L1482 12CO 13CO C18O dn=60 VGT vs Planck}
\end{figure*}

\begin{figure*}
	\includegraphics[width=0.99\linewidth]{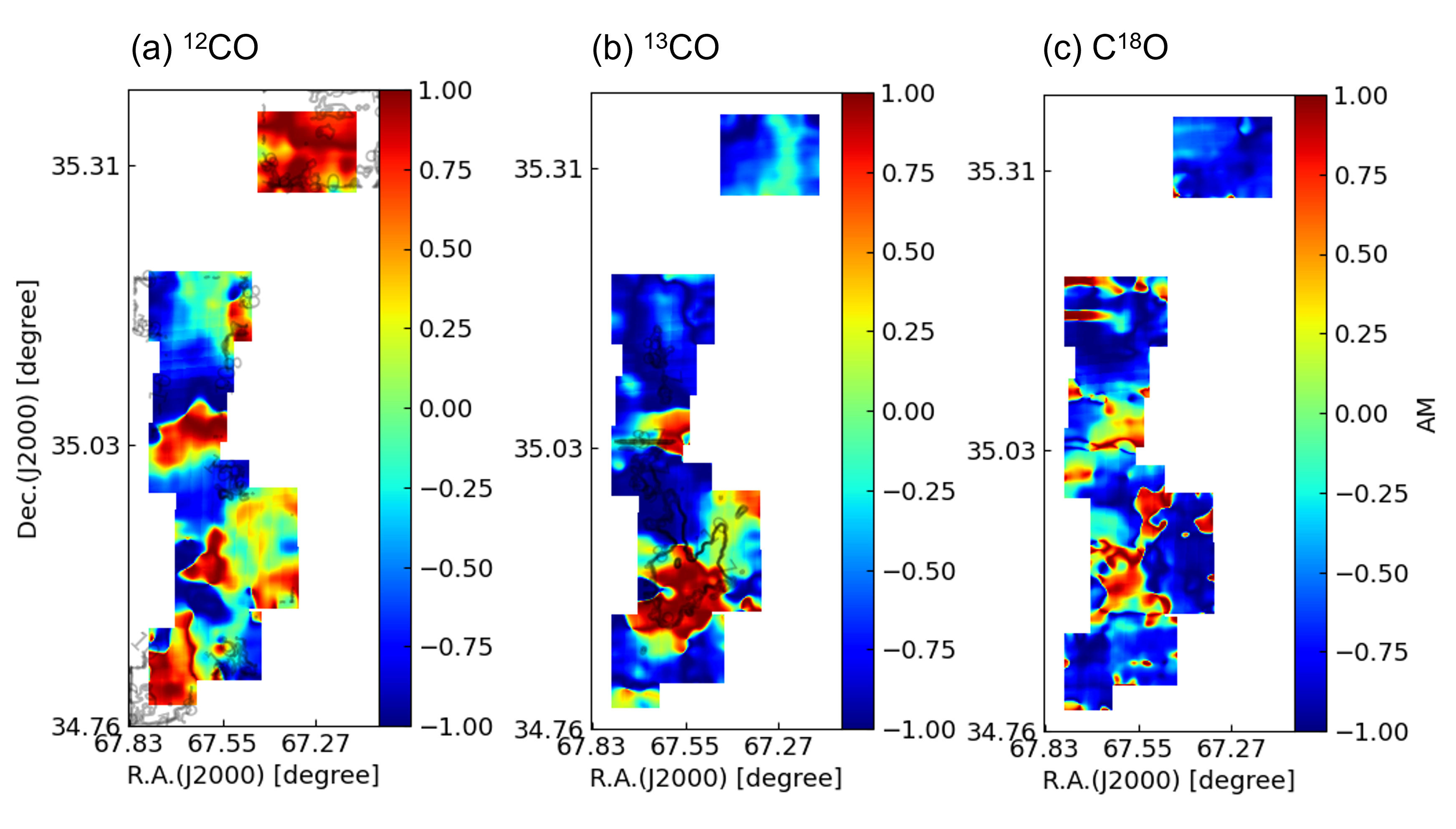}
    \caption{The AM inferred from the measurement of alignment of the Planck and VGT gradients using $^{12}$CO (left), $^{13}$CO (middle), and C$^{18}$O (right), respectively, for sub block sizes of 60$\times$60 pixels ($\sim1.2$~pc) in the L1482 region of the CMC. Red regions represent parallel alignment and AM values of one and blue regions represent perpendicular alignment with AM value of negative one.}
    \label{fig:L1482 12CO 13CO C18O dn=60 AM map}
\end{figure*}

\begin{figure}
	\includegraphics[width=1.0\linewidth]{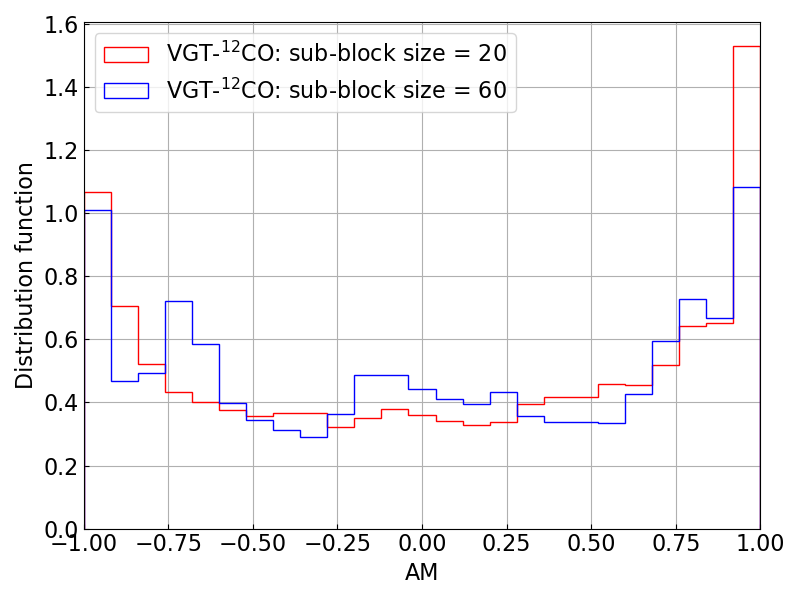}
    \caption{Same as Fig.~\ref{fig:L1478 12CO Histogram}, but for the L1482 region of the CMC.}
    \label{fig:L1482 12CO Histogram}
\end{figure}

\begin{figure}
	\includegraphics[width=1.0\linewidth]{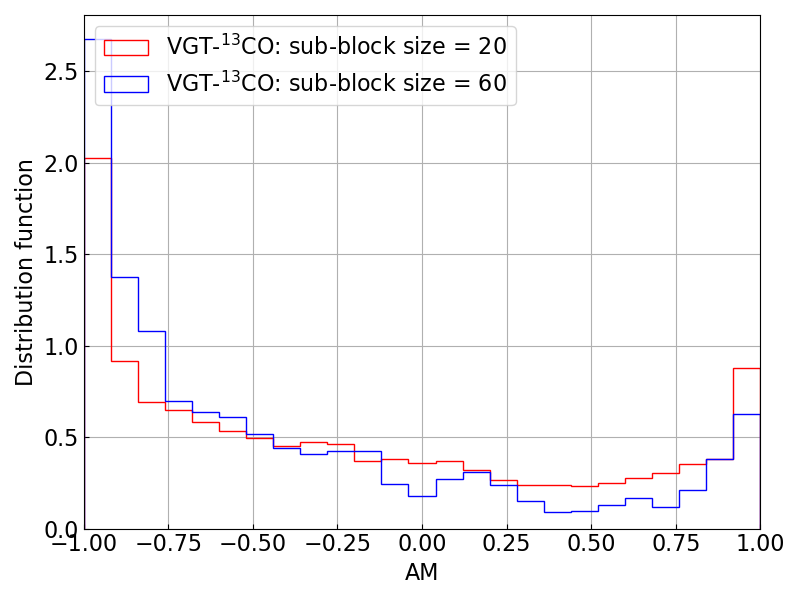}
    \caption{Same as Fig.~\ref{fig:L1482 12CO Histogram}, but for $^{13}$CO.}
    \label{fig:L1482 13CO Histogram}
\end{figure}

\begin{figure}
	\includegraphics[width=1.0\linewidth]{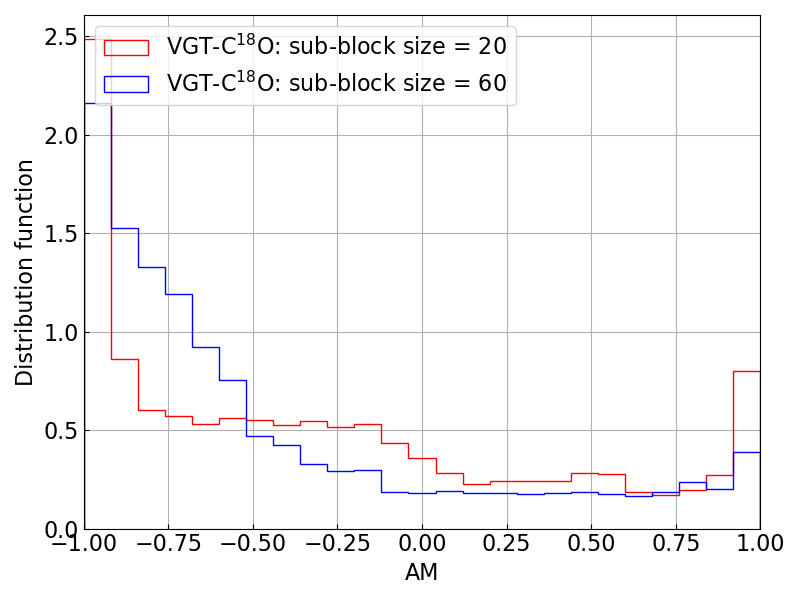}
    \caption{Same as Fig.~\ref{fig:L1482 12CO Histogram}, but for C$^{18}$O.}
    \label{fig:L1482 C18O Histogram}
\end{figure}

\subsection{L1482}
L1482 is another sub-cloud of the CMC. It is located to the east of L1478, however, it shows higher column density \citep{2017A&A...606A.100L,2018A&A...620A.163Z}. The high column density suggests stronger self-gravity in L1482. As we discussed in \S~\ref{sec:method}, strong self-gravity could change the velocity gradient's direction from perpendicular to parallel to the magnetic field. In terms of AM, such a change results in negative values. To exclude the difference in resolution size's effect on VGT, we chose to mainly use the $60\times60$ sub-block sized (i.e., 10$'$) maps instead of 3.3$'$ maps in order to examine the self-gravity's effect on VGT and exclude the contribution from resolution differences.

The comparison of VGT and Planck, as well as the spatial distribution of AM, is presented in Fig.~\ref{fig:L1482 12CO 13CO C18O dn=60 VGT vs Planck} and \ref{fig:L1482 12CO 13CO C18O dn=60 AM map}. Considering the optically thick tracer $^{12}$CO, L1482 appears to have a mix of perpendicular alignment and parallel alignment spread out in clumps. Particularly the northern small region shows great alignment with AM larger than 0.60. The southern majority of the clump, however, exhibits more negative AM. The mean AM values for 3.3$'$ and 10$'$ are 0.06 and 0.02 respectively (see Tab.~\ref{tab:example_table}). These values are much lower than the mean AM values of L1478 when using $^{12}$CO as a tracer. Looking at denser tracer $^{13}$CO, much more negative-AM regions appear in Fig.~\ref{fig:L1482 12CO 13CO C18O dn=60 AM map} suggesting perpendicular alignment of velocity gradient and polarization. For $^{13}$CO the mean AM values with 3.3$'$ and 10$'$ resolution are -0.22 and -0.40, respectively. These values are much more negative than our mean AM values for $^{12}$CO. This means there is much more perpendicular alignment at $^{13}$CO. The most apparent change is observed in the northern small region. The positive AM $>0.6$ for VGT-$^{12}$CO turns into negative. Such change means that self-gravity starts dominating at volume density $>10^3~{\rm cm^{-3}}$ which is the density range traced by $^{13}$CO. Importantly, we find the negative AM typically appears in ambient low-intensity region, while the high-density clump exhibits positive AM close to 1, which is even higher than that of VGT-$^{12}$CO. This central clump corresponds to the location of the massive stellar cluster NGC1579 \citep{2014A&A...567A..10L} associated with a H II expanding bubble and stellar wind \citep{2004AJ....128.1233H}. The H II region may have more significant effects on changing diffuse molecular tracer $^{12}$CO's dynamics so that VGT-$^{12}$CO shows smaller AM than VGT-$^{13}$CO.


This negative AM, however, is less likely to be contributed by the foreground or background because we do not expect a perpendicular alignment between velocity gradient and foreground or background magnetic fields. The possible explanation is that L1482 is under global gravitational contraction so that it is acreting ambient gas into the central dense clump. Stronger magnetic field and turbulence may counteract the collapse at the clump. This interesting phenomenon requires further studies. 

As for the densest tracer C$^{18}$O used in this work, we don't see as much of a drastic change in perpendicular alignment, as we did when comparing $^{13}$CO to $^{12}$CO but there is still a minor magnitude increase in negative values. The mean AM values measured for C$^{18}$O with 3.3$'$ and 10$'$ resolution\footnote{We present only the gradient vector maps at 10' to avoid the confusion of resolution difference and self-gravity effect. However, we conducted the calculation for both 3.3’ and 10’ and we list the AM value in Tab.1.} are -0.29 and -0.45. This confirms our explanation that L1482 could be experiencing gravitational contraction at volume density $>10^3~{\rm cm^{-3}}$ so that we observed similar AM distribution for the denser tracers C$^{18}$O.

Furthermore, Figs.~\ref{fig:L1482 12CO Histogram}, \ref{fig:L1482 13CO Histogram}, and \ref{fig:L1482 C18O Histogram} present the histograms of the AM at two resolutions 10$'$ and 3.3$'$. For VGT-$^{12}$CO's histogram, we see two peaks appears in ${\rm AM} = 1$ and ${\rm AM}=-1$ suggesting the self-gravity's effect occurs. As for VGT-$^{13}$CO and VGT-C$^{18}$O, the most apparent peak appears only on negative ${\rm AM}=-1$. Particularly, the histogram at high resolution 3.3$'$ is less dispersed than the 10$'$ one due to the difference in VGT and Planck's resolution. Nevertheless, the globally negative AM means self-gravity's effect is significant in L1482.

\section{Discussion}
\label{sec:dis}
\subsection{The physical properties of L1478 and L1482}
L1478 and L1482 are two sub-cloud of the giant CMC. Even though the two are closely located, they show very different physical properties. In our comparison of VGT and Planck polarization, we find in L1478 the velocity gradients are parallel to the magnetic field inferred from polarization, while in L1482 their relative orientation is mostly perpendicular. 

According to earlier studies \citep{HLY20}, the parallel alignment suggests that self-gravity dominates over turbulence. It means L1482 at scales of order $\sim1~{\rm pc}$ undergoes global gravitational contraction. The ambient gas is accreting into the clump's center so that the velocity gradient is dominated by the contribution from the inflow. This agrees with the earlier study by \cite{2014A&A...567A..10L} suggesting that L1482 is in the process of gravitational fragmentation exhibiting  converging inflows. With VGT, we get more fruitful information. Such perpendicular alignment in L1482 is partially observed in VGT-$^{12}$CO and is more apparent in VGT-$^{13}$CO and VGT-C$^{18}$O. Considering that $^{12}$CO, $^{13}$CO, and C$^{18}$O typically trace molecular gas at volume density ranges $\sim10^{2}~{\rm cm^{-3}}$, $10^{3}~{\rm cm^{-3}}$, and $10^{4}~{\rm cm^{-3}}$, the significance in VGT-$^{13}$CO and VGT-C$^{18}$O reveals that the gravitational contraction mainly happens at volume density $>10^{3}~{\rm cm^{-3}}$.  

The situation in L1478 gets different. The parallel alignment between the velocity gradients and the magnetic field inferred from polarization means turbulence is dominated and self-gravity's effect is negligible. Particularly, we test the VGT at two resolutions 3.3$'$ and 10$'$. We see that the alignment increases when comparing VGT and Planck at the same resolution 10$'$. The difference observed in VGT and polarization measurements can be contributed by the difference in resolution. This importantly highlights the VGT's advantage in tracing the magnetic field at smaller scales using emission lines.

\subsection{Uncertainties}
\label{sec:unc}
In this study, we investigate the L1478 and L1482 clouds by comparing VGT and Planck polarizations. We acknowledge several potential uncertainties in the comparison. Firstly, both VGT and Planck have inherent uncertainties. The uncertainty in VGT mainly stems from the sub-block averaging method and the noise in spectroscopic observations. To address these factors, we calculated the uncertainty maps of VGT, as demonstrated in Figs.~\ref{fig:L1478 Uncertainty Maps} and \ref{fig:L1482 Uncertainty Maps}. Secondly, dust polarization and VGT rely on distinct physical mechanisms to trace magnetic fields, which may lead to misaligned magnetic field measurements between Planck and VGT. For example, dust polarization includes foreground/background emissions, whereas VGT directly probes the magnetic field associated with molecular clouds. Additionally, VGT uses molecular tracers $^{12}$CO, $^{13}$CO, and C$^{18}$O, which are sensitive to different critical density values. The polarization based on RATs-aligned dust provides LOS integrated measurements, and the magnetic fields measured may be inhomogeneous if RATs are too strong in high-density regions. Despite these uncertainties, the overall statistically significant parallel (in L1478) and perpendicular (in L1482) alignments of VGT and Planck suggest that these factors are sub-dominated.

\section{Conclusion}
\label{sec:con}
In this work, we use the velocity gradient technique (VGT) to trace the POS magnetic field orientation in the CMC's two sub-clouds L1478 and L1482, and determine the dominance of MHD turbulence or self-gravity. We apply VGT to three emission lines $^{12}$CO, $^{13}$CO, and C$^{18}$O to get information over the volume density range in approximately from 10$^{2}$~cm$^{-3}$ to 10$^{4}$~cm$^{-3}$. We compare VGT results calculated in the resolutions of $3.3'$ and $10'$ to Planck polarization at 353 GHz and $10'$ to determine areas of MHD turbulence dominance and gravitational collapse dominance. Our main findings are:
\begin{enumerate}    
    \item We found the VGT-measured magnetic fields globally agree with that from Planck in L1478 suggesting self-gravity's effect is insignificant. The best agreement appears in VGT-$^{12}$CO.
    \item For the turbulence-dominated cloud L1478, we use VGT with $^{12}$CO, $^{13}$CO, and C$^{18}$O to provide magnetic field orientation maps at resolution $\sim3.3'$, higher than Planck polarization.
    \item We show that the resolution difference can introduce misalignment between VGT and Planck polarization measurement.
    \item As for L1482, the VGT measurements are statistically perpendicular to the Planck polarization indicating the dominance of self-gravity. This perpendicular alignment is more significant in  VGT-$^{13}$CO and VGT-C$^{18}$O.
\end{enumerate}


\section*{Acknowledgements}
Y.H. and A.L.acknowledges the support of NASA ATP AAH7546. Financial support for this work was provided by NASA through award 09\_0231 issued by the Universities Space Research Association, Inc. (USRA).
\section*{Data Availability}
The data underlying this article will be shared on reasonable request to the corresponding author.


\bibliographystyle{mnras}
\bibliography{example} 




\appendix
\section{Uncertainty Maps}
\label{sec:appA}
\begin{figure*}
	\includegraphics[width=1.0\linewidth]{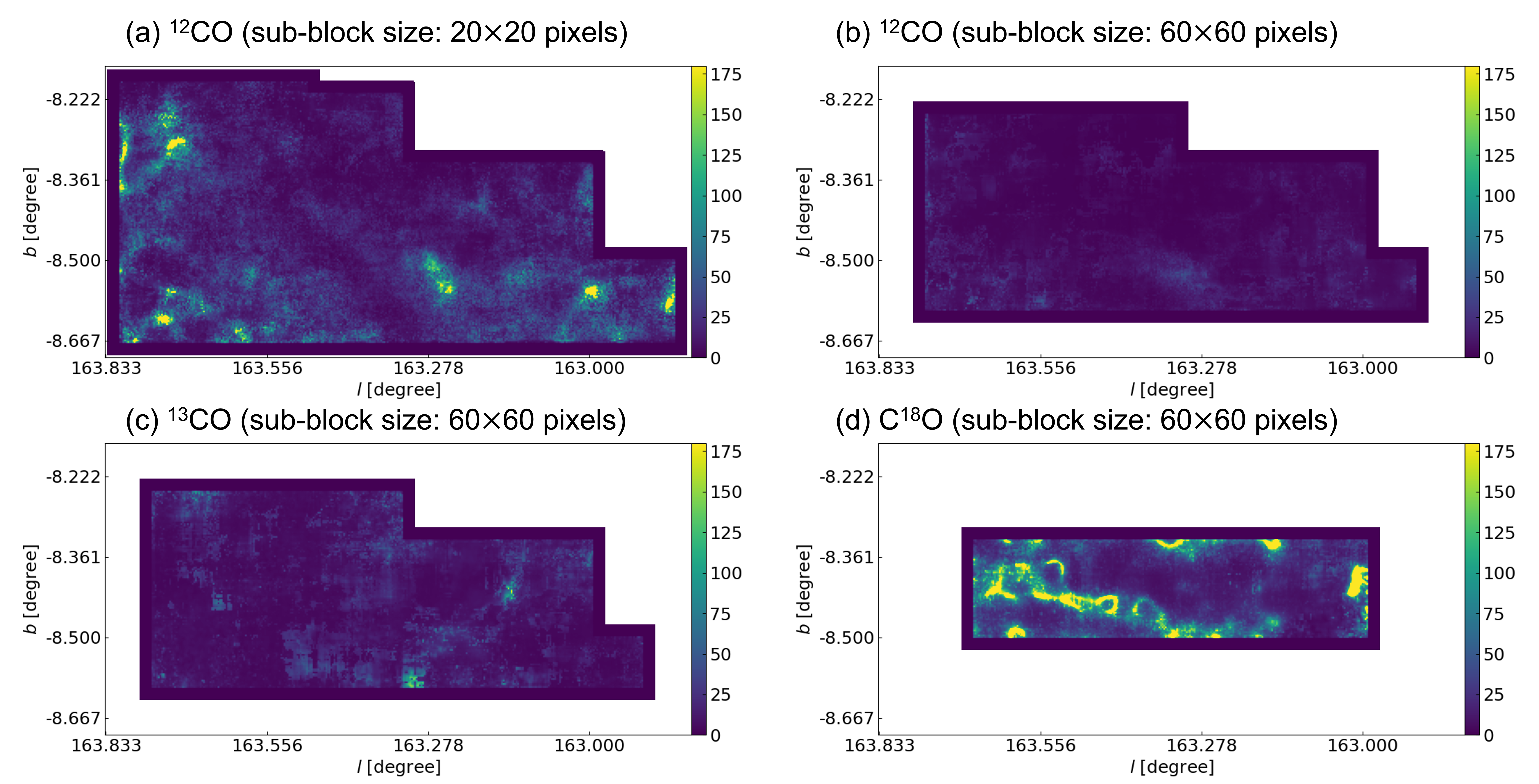}
    \caption{Uncertainty maps for L1478. The color bar is in the unit of degrees with range 0 - 180. 
    }
    \label{fig:L1478 Uncertainty Maps}
\end{figure*}

\begin{figure*}
	\includegraphics[width=0.8\linewidth]{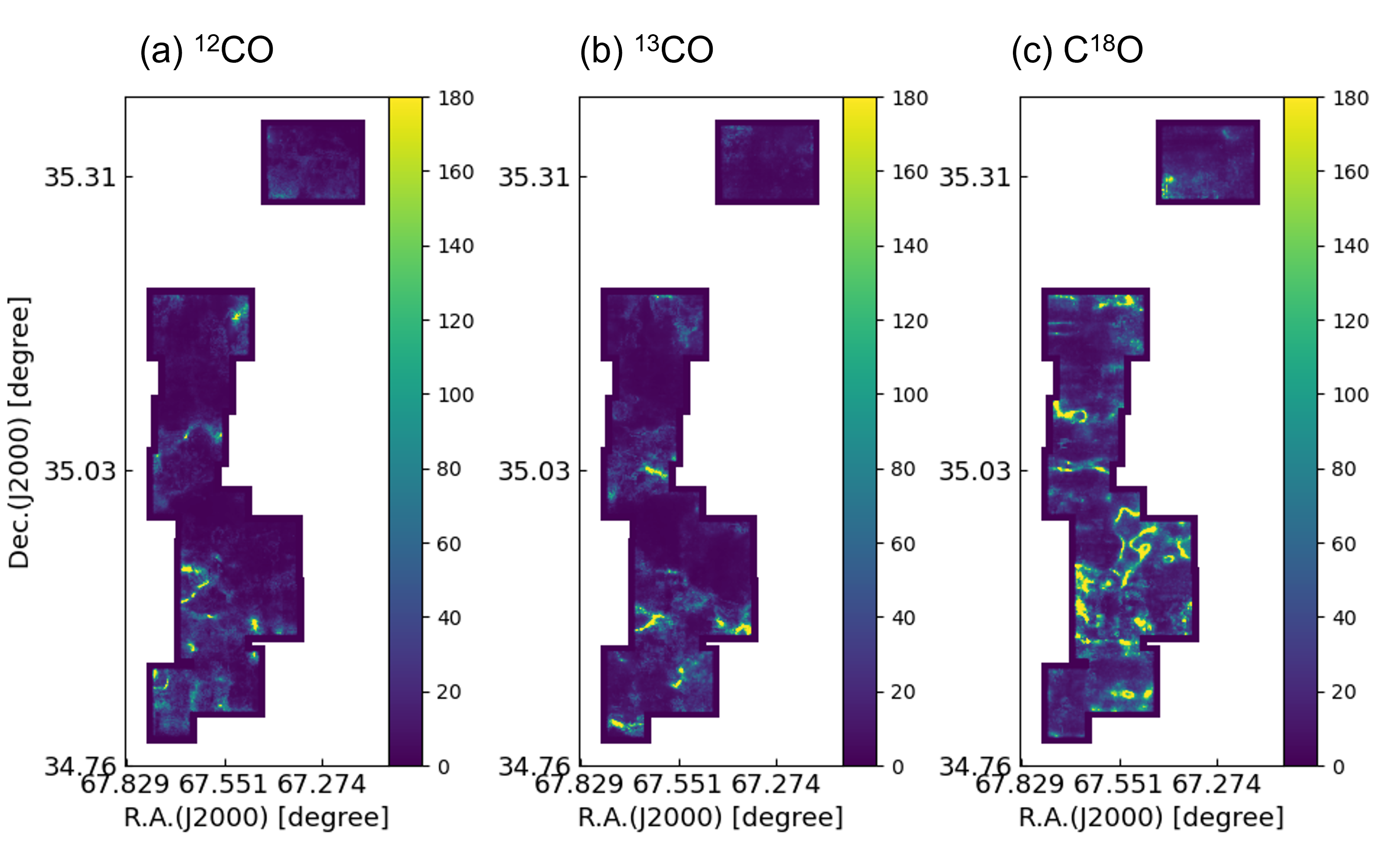}
    \caption{Uncertainty maps for L1482. The color bar is in the unit of degrees with range 0 - 180.
    }
    \label{fig:L1482 Uncertainty Maps}
\end{figure*}

Figs.~\ref{fig:L1478 Uncertainty Maps} and \ref{fig:L1482 Uncertainty Maps} present the uncertainty maps of our VGT measurements. Such uncertainties can be mostly attributed to the systematic error from the observational data as well as the VGT algorithm. As introduced in Sec.~\ref{sec:method}, the algorithm fits the gradient’s orientations over predetermined sub-regions into a Gaussian histogram and yields the angle at which exists the peak value of the histogram. This error from the Gaussian fitting algorithm within the 95\% confidence level is denoted as $\sigma_{\psi_{gs}} (x, y, v)$, and thus the uncertainties are obtained as such:
\begin{equation}
\begin{aligned}
    &\sigma_{{\rm cos}} (x,y,v)  =  |2 {\rm sin}(2 \psi_{\rm gs} (x,y,z)) \sigma_{{\psi_{\rm gs}}} (x,y,v)|\\
    &\sigma_{{\rm sin}} (x,y,v)  =  |2 {\rm cos}(2 \psi_{\rm gs} (x,y,z)) \sigma_{{\psi_{\rm gs}}} (x,y,v)|\\
    &\sigma_{q} (x,y,v)  =  |{\rm Ch} \cdot {\rm cos(2 \psi_{\rm gs})}|\sqrt{(\sigma_{n}/{\rm Ch})^2+(\sigma_{{\rm cos}}/{\rm cos(2 \psi_{\rm g})})^2}\\
    &\sigma_{u} (x,y,v)  =  |{\rm Ch} \cdot {\rm sin(2 \psi_{\rm gs})}|\sqrt{(\sigma_{n}/{\rm Ch})^2+(\sigma_{{\rm sin}}/{\rm sin(2 \psi_{\rm g})})^2}\\
    &\sigma_{Q} (x,y)  =  \sqrt{\sum_{v} {\sigma_{q}(x,y,v)^2}}\\
    &\sigma_{U} (x,y)  =  \sqrt{\sum_{v} {\sigma_{u}(x,y,v)^2}}\\
    &\sigma_{\psi_{g}} (x,y) = \frac{|U_{\rm g}/Q_{\rm g}|\sqrt{(\sigma_{Q}/Q_g)^2+(\sigma_{U}/U_g)^2}}{2[1+(U_{\rm g}/Q_{\rm g})^2]}
\end{aligned}
\end{equation}
where $\sigma_{\psi_g}$ is the angular uncertainty of the magnetic field measurements, $\sigma_n (x, y, v)$ denotes the noise in velocity channel Ch$(x, y, v)$ as well as error propagation, and $\sigma_Q (x, y)$, $\sigma_U(x, y)$ give the respective uncertainty of the pseudo-Stokes parameters $Q_g (x, y)$, $U_g (x, y)$. 

\section{Line spectra}
\begin{figure*}
	\includegraphics[width=0.8\linewidth]{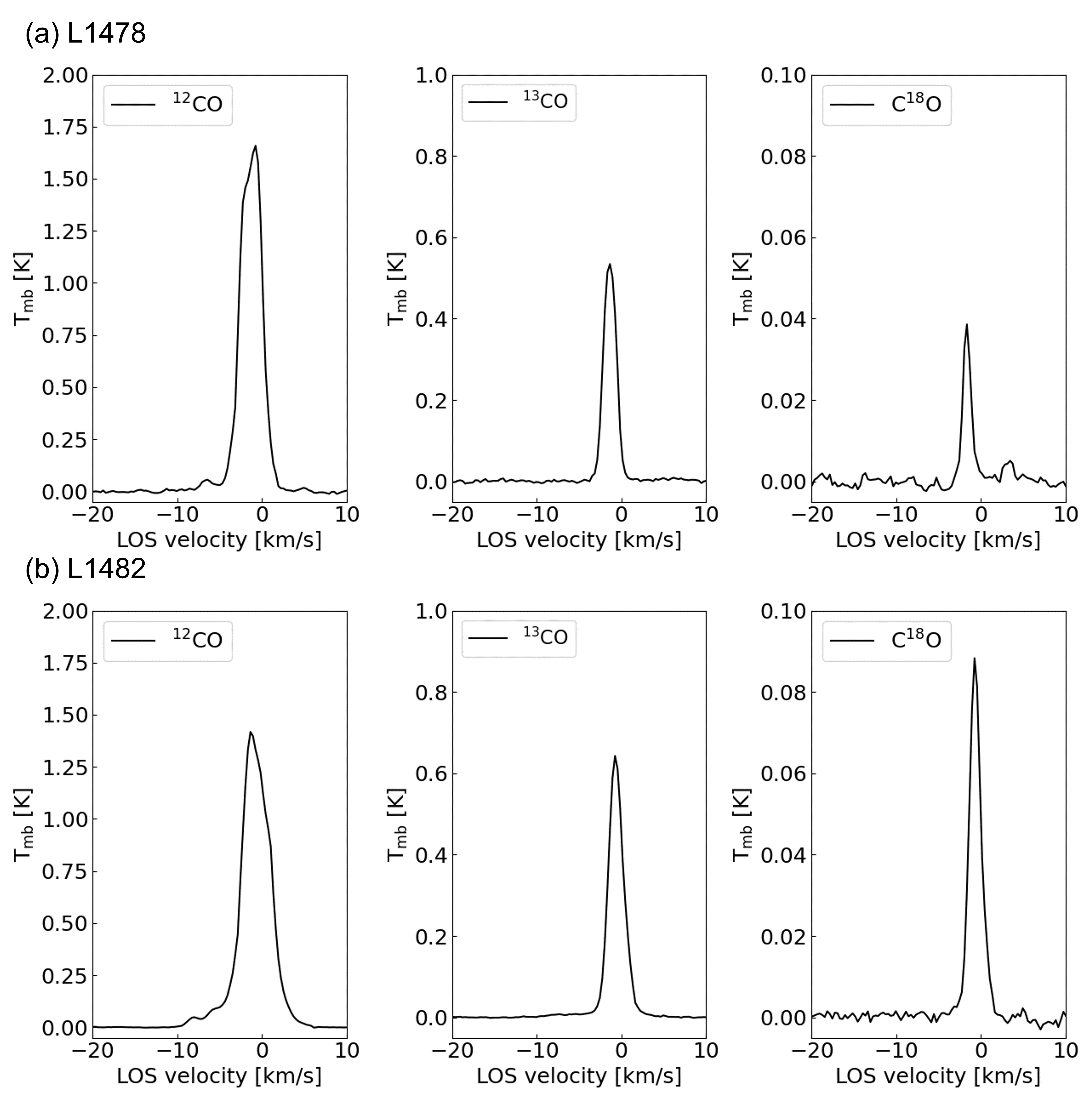}
    \caption{Average line spectra for L1478 and L1482.}
    \label{fig:L1478 and L1482 Spectra}
\end{figure*}

Fig.~\ref{fig:L1478 and L1482 Spectra} presents the averaged line spectra for L1478 and L1482. In L1478, the $^{12}$CO line exhibits a maximum intensity of approximately 1.7 K, while the $^{13}$CO line shows a lower intensity with a maximum of around 0.55 K. The C$^{18}$O line has an even lower intensity with a maximum of about 0.04 K. All three spectra have different noise levels, with $^{12}$CO having a higher noise level of below 0.05 K compared to $^{13}$CO and C$^{18}$O, which have noise levels below 0.02 K and 0.005 K, respectively. In addition, for L1482, the $^{12}$CO line exhibits a higher maximum intensity of approximately 1.4 K, while the $^{13}$CO and C$^{18}$O lines have lower intensities with maximum values of around 0.65 K and 0.09 K, respectively. The noise levels for the three lines also differ, with $^{12}$CO having a higher noise level of below 0.1 K, $^{13}$CO having a noise level below 0.03 K, and C$^{18}$O having a noise level below 0.005 K. Both L1487 and L1482 are dominated by a single velocity component.

\section{Sub-block averaging}
\begin{figure*}
	\includegraphics[width=0.6\linewidth]{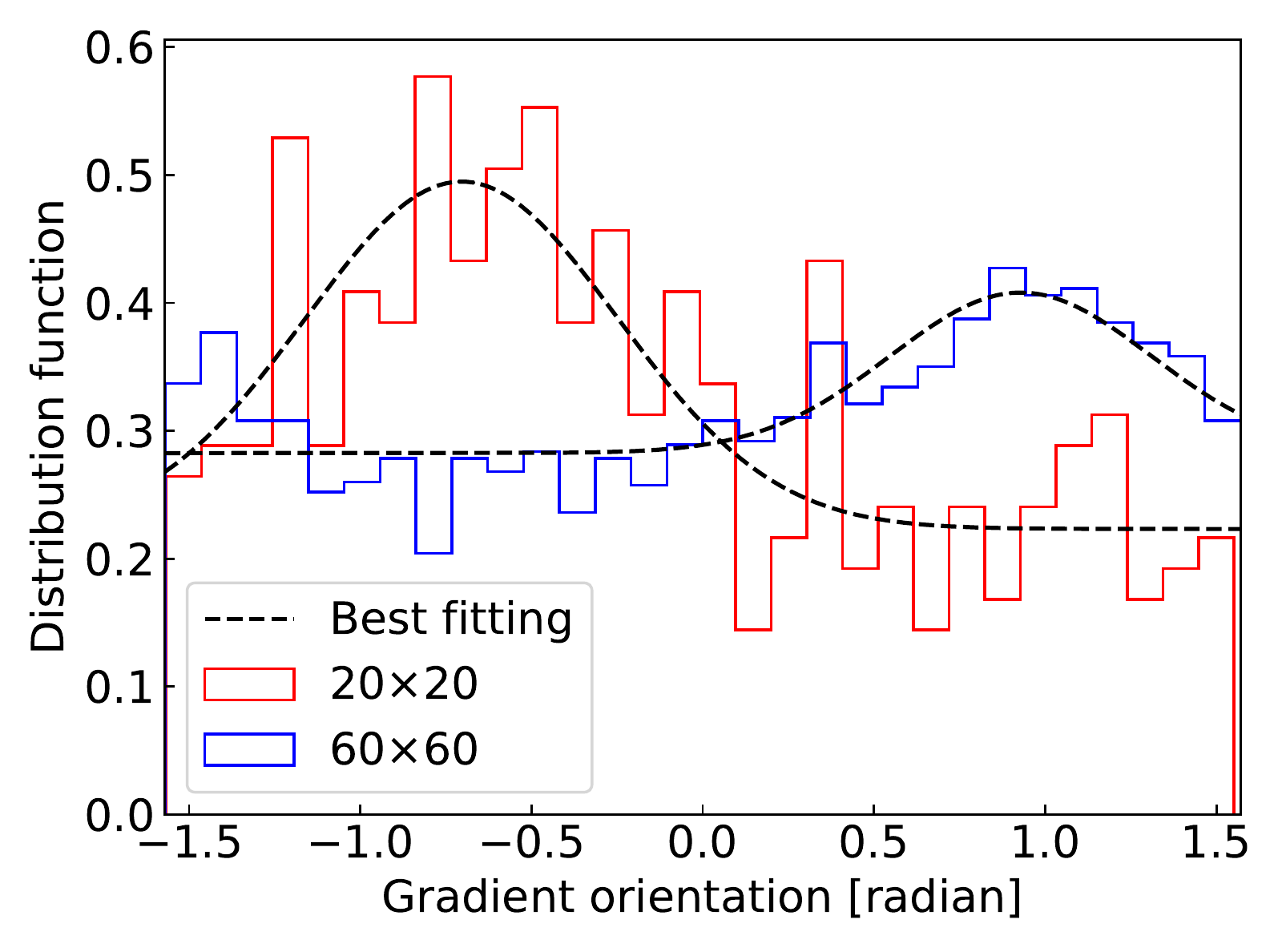}
    \caption{Histogram of gradient orientation for two randomly selected sub-blocks within L1478. }
    \label{fig:subblock}
\end{figure*}

The sub-block averaging method implemented in VGT requires that the sampled gradients form a Gaussian distribution. Here we present a test for this requirement. In Fig.~\ref{fig:subblock}, we plot the histogram of gradient orientation for two randomly selected sub-blocks within L1478. The two sub-blocks have sizes of $20\times20$ pixels and $60\times60$ pixels, respectively. The histogram exhibits two apparent Gaussian distributions.

\bsp	
\label{lastpage}
\end{document}